\documentclass[apjl]{emulateapj}

\newcommand{\jms}{J. Mol. Spec.}
\newcommand{\jmst}{J. Mol. Struct.}
\newcommand{\pss}{Plan. Sp. Sc.}

\newcommand{\aapl}{\aap~Letters}
\newcommand{\aspc}{ASP Conf. Series}
\renewcommand{\jqsrt}{J. Quant. Spec. Radiat. Transf.}
\newcommand{\subscript}[1]{\textnormal{\scriptsize{#1}}}
\newcommand{\rout}{\ensuremath{R_{d3}}}
\newcommand{\rmida}{\ensuremath{R_{d1}}}
\newcommand{\rmidb}{\ensuremath{R_{d2}}}
\newcommand{\rdin}{\ensuremath{R_c}}

\newcommand{\tdin}{\ensuremath{T_d(\rdin)}}
\newcommand{\tdout}{\ensuremath{T_d(\rout)}}

\newcommand{\tvib}{\ensuremath{T_\subscript{vib}}}
\newcommand{\trot}{\ensuremath{T_\subscript{rot}}}
\newcommand{\acet}{\ensuremath{\textnormal{C}_2\textnormal{H}_2}}
\newcommand{\isoacet}{\ensuremath{\textnormal{H}^{13}\textnormal{CCH}}}
\newcommand{\diacet}{\ensuremath{\textnormal{C}_4\textnormal{H}_2}}
\newcommand{\isodiacet}{\ensuremath{\textnormal{H}^{13}\textnormal{CCCCH}}}
\newcommand{\isodosdiacet}{\ensuremath{\textnormal{HC}^{13}\textnormal{CCCH}}}
\newcommand{\triacet}{\ensuremath{\textnormal{C}_6\textnormal{H}_2}}
\newcommand{\tetracet}{\ensuremath{\textnormal{C}_8\textnormal{H}_2}}
\newcommand{\hcn}{\ensuremath{\textnormal{HCN}}}
\newcommand{\isohcn}{\ensuremath{\textnormal{H}^{13}\textnormal{CN}}}
\newcommand{\hcccn}{\ensuremath{\textnormal{HC}_3\textnormal{N}}}
\newcommand{\kms}{km~s$^{-1}$}
\newcommand{\cm}{cm$^{-1}$}
\newcommand{\cmatm}{cm$^{-2}$~atm$^{-1}$}
\newcommand{\science}{Science}
\newcommand{\icarus}{Icarus}
\newcommand{\nature}{Nature}

\begin{document}

\title{The abundances of polyacetylenes towards CRL618}
\shorttitle{The abundances of polyacetylenes towards CRL618}
\shortauthors{J. P. Fonfr\'{\i}a et al.}
\author{J. P. Fonfr\'{\i}a\altaffilmark{1}}
\author{J. Cernicharo\altaffilmark{2}}
\author{M. J. Richter\altaffilmark{3,5}}
\author{J. H. Lacy\altaffilmark{4,5}}
\affiliation{
$^1$Depto. de Estrellas y Medio Interestelar, Instituto de Astronom\'ia, 
UNAM, Ciudad Universitaria, 04510, Mexico City (Mexico)\\
$^2$Laboratorio de Astrof\'isica Molecular, Dpto. de Astrof\'isica,
Centro de Astrobiolog\'ia, INTA-CSIC, 28850
Torrej\'on de Ardoz, Madrid (Spain)\\
$^3$Physics Dept. - UC Davis, One Shields Ave., Davis,
CA 95616 (USA)\\
$^4$Astronomy Dept., University of Texas, Austin,
TX 78712 (USA)}
\altaffiltext{5}{Visiting Astronomer at the Infrared Telescope Facility,
which is operated by the University of Hawaii under
contract from the National Aeronautics and Space Administration.}

\begin{abstract}
We present a mid-infrared high spectral resolution spectrum of CRL618
in the frequency ranges $778-784$ and $1227-1249$~\cm{}
($8.01-8.15$ and $12.75-12.85~\mu$m)
taken with the Texas Echelon-cross-Echelle Spectrograph (TEXES)
and the Infrared Telescope Facility (IRTF).
We have identified more than 170 ro-vibrational lines
arising from \acet{}, \hcn{}, \diacet{}, and \triacet{}.
We have found no unmistakable trace of \tetracet{}.
The line profiles display a complex structure suggesting 
the presence of 
polyacetylenes
in several components of the circumstellar envelope (CSE).
We derive total column densities of $2.5\times 10^{17}$, 
$3.1\times 10^{17}$, $2.1\times 10^{17}$, $9.3\times 10^{16}$~cm$^{-2}$, and 
$\lesssim 5\times 10^{16}$~cm$^{-2}$ for \hcn{}, \acet{}, \diacet{}, 
\triacet{}, and \tetracet, respectively. 
The observations indicate that both the rotational and vibrational 
temperatures in the innermost CSE depend 
on the molecule, varying from $100$ to $350$~K for the rotational 
temperatures and $100$ to $500$~K for the vibrational temperatures.
Our results support a chemistry in the innermost 
CSE based on radical-neutral reactions triggered by the intense 
UV radiation field.
\end{abstract}

\maketitle

\section{Introduction}
\label{sec:intro}

The protoplanetary nebula stage (PPN) is one of the shortest
of a sun-like star's evolution \citep*[e.g.,][]{iben_1983}.
Throughout this phase, roughly half of the stellar photosphere
is ejected, shocking the gas of the circumstellar envelope (CSE)
formed in the AGB stage \citep*[AGB-CSE;][]{kwok_1978}, 
and unveiling the outermost layers
of the nucleus.
The stellar UV radiation field is extremely intense in this phase, 
photodissociating
the innermost circumstellar gas and triggering a particularly rich 
photochemistry
\citep*[][hereafter C04]{woods_2002,woods_2003,redman_2003,cernicharo_2004}.

CRL618 (Westbrook Nebula, AFGL 618) is a very young PPN 
\citep*[age $\simeq 200$~yr;][]{kwok_1984},
located at a distance of $\simeq 0.9-1.8$~kpc from the Sun
\citep{schmidt_1981,goodrich_1991,knapp_1993,sanchezcontreras_2004a}.
It contains a B0 central star embedded in a dusty ultracompact HII region
surrounded by a torus and a low-velocity expanding envelope
with an external radius $>20$\arcsec, 
a total mass $\simeq 1$~M$_\odot$, and an expansion velocity
$v_\subscript{exp}\simeq 18.0-21.5$~\kms{}
\citep{knapp_1985,fuente_1998,pardo_2004}.
In addition, it displays gas with velocities as high as 200~\kms{}
\citep{burton_1986,cernicharo_1989,gammie_1989}.
This high velocity gas (HVG) is the molecular counterpart of the bright
optical jets oriented in the E-W direction \citep{trammell_2000} which
impact the AGB-CSE and produce the well-known optical lobes.

Since its discovery by \citet{westbrook_1975} many molecular species
have been detected towards this PPN
\citep{cernicharo_1989, cernicharo_2001a, cernicharo_2001b,
herpin_2000,remijan_2005,truong-bach_1996}, some of them for 
the first time in a C-rich CSE
\citep*[e.g., formaldehyde, polyacetylenes \diacet{} and \triacet{},
benzene, H$_2$O and OH;][]{cernicharo_1989,cernicharo_2001a,
cernicharo_2001b,herpin_2000}.
The CSE developed in the AGB phase has been studied in great detail by several
authors 
\citep*[see, e.g.,][and references therein]{cernicharo_2001a,
cernicharo_2001b, sanchezcontreras_2004a,
sanchezcontreras_2004b,pardo_2004,pardo_2005,pardo_2007a,pardo_2007b}.

\citet*[][hereafter W03]{woods_2003} and C04 have modelled the 
chemistry of CRL618 suggesting that in the 
innermost envelope the UV photons photodissociate \acet{} producing 
the radical C$_2$H, which can react with \acet{} to form \diacet{} or 
with H$_2$ reforming \acet{}.
Additionally, C$_2$H can react with \diacet{} forming \triacet{}.
These processes lead to a rapid \acet{} polymerization 
in long carbon chains and clusters.
The abundance ratio between consecutive polyacetylenes (C$_{2n}$H$_2$,
$n=1,2,3,\ldots$) in CRL618 has been estimated as
a factor $\simeq 2-3$ \citep*[][hereafter C01a]{cernicharo_2001a}.
Polyacetylenes are symmetric molecules without a permanent
dipole moment and, hence, detectable only through their ro-vibrational
spectra.
The strongest bands of their spectra are expected to fall in the
mid-infrared range due to the physical conditions prevailing in the
innermost CSE where the polyacetylenes are built up.
However, the large telluric opacity in the infrared
has largely prevented the exploitation of this frequency range.
This issue has been overcome with the launching of the 
Infrared Space Observatory (ISO) and the Spitzer Space Telescope, 
and the developing of
instruments with high spectral resolving power such as
the Texas Echelon-cross-Echelle Spectrograph \citep*[TEXES;][]{lacy_2002}.
The analysis of these kinds of observations will allow us to improve
our knowledge about the polymerization processes and the formation of
complex molecules such as long carbon chains, PAHs, and fullerenes 
(C$_{60}$ and C$_{70}$), which have been recently observed
towards circumstellar and interstellar environments
\citep{cami_2010,garciahernandez_2010,sellgren_2010} and whose ions could be the carriers of the Diffuse 
Interstellar Bands \citep*[DIBs;][]{foing_1994}.

In this paper, we present a high resolution mid-infrared spectrum 
of CRL618.
We have identified $\simeq 170$ ro-vibrational lines of bands
$\nu_6+\nu_8$, $\nu_6+\nu_8+\nu_9-\nu_9$, $\nu_6+\nu_8+2\nu_9-2\nu_9$,
$\nu_6+\nu_8+\nu_7-\nu_7$ of \diacet{}, and $\nu_8+\nu_{11}$ of \triacet{}.
In addition, we have observed several lines of \acet{} and \hcn{}.
These lines have been analyzed by using a modified version of the model of AGB 
envelopes developed by \citet*[][hereafter F08]{fonfria_2008}.
The observations and the spectroscopic laboratory data 
of \hcn{} and polyacetylenes required to analyze the observations
are presented in 
\S\S\ref{sec:observations} and \ref{sec:spectroscopy}, respectively.
A description of the spectrum and the model, the adopted
fitting strategy, and a discussion about the uncertainties
of the parameters can be found in \S\ref{sec:modeling}.
The results derived from our fits are presented and 
discussed in \S\ref{sec:results} and,
finally, summarized in \S\ref{sec:conclusions}.

\section{Observations}
\label{sec:observations}

\begin{figure*}
\centering
\includegraphics[angle=0,height=0.9\textheight]{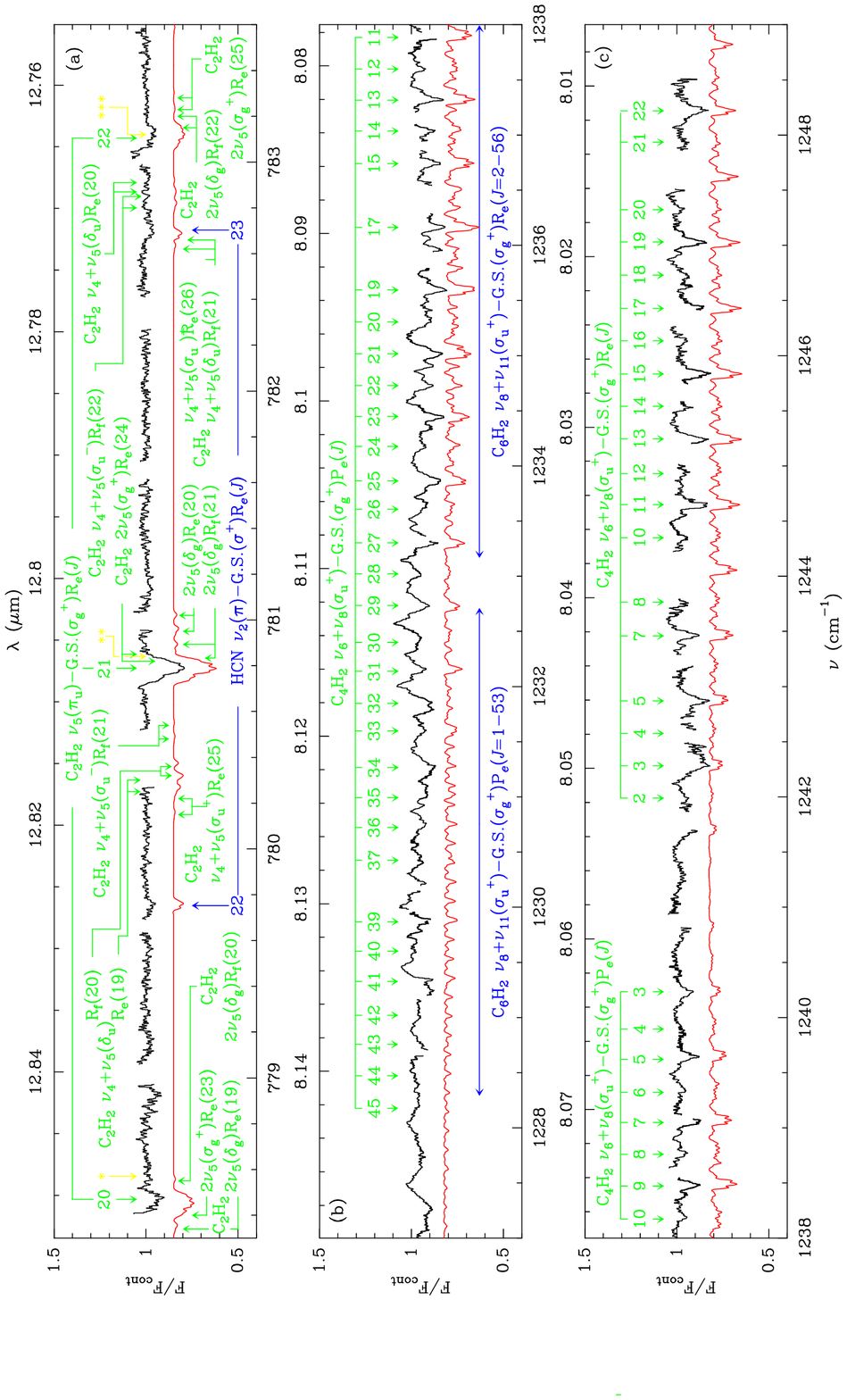}
\caption{Observed and synthetic spectra towards CRL618 (black and red).
($a$) Spectra in the range $778-784$~\cm{}.
The detected lines are produced by \acet{} and \hcn{} (green and blue).
The asterisks (yellow) indicate several absorptions that
could be assigned to the \isoacet{} lines
$\nu_5\textnormal R_e(21)$, $\textnormal R_e(22)$, and $\textnormal R_e(23)$
(*, **, and ***, respectively).
($b$ and $c$) Spectra ranging from 1227 to 1249~\cm{} contain
\diacet{} $\nu_6+\nu_8$ and \triacet{} $\nu_8+\nu_{11}$ bands (green and blue).
See the text for the modeling details of these lines.
[\textit{See the electronic edition of the Journal for a color version of
this figure.}]}\label{fig:f1}
\end{figure*}

We observed CRL618 using the 
TEXES spectrograph
\citep{lacy_2002} on the IRTF 3~m telescope during the first
half of the nights of 12 Dec 2002 and 8 Jan 2004. TEXES was used in 
high spectral resolution mode which gives $R\simeq 10^5$. Six separate settings
were used to cover from 778 to 784~\cm{} (2002/12/12) and
1205 to 1250~\cm{} (2004/01/08; see Fig.~\ref{fig:f1}).
For each setting, the Becklin-Neugebauer Object 
\citep*[BN;][]{becklin_1967} was observed as a telluric reference before 
changing the instrumental setup. To date, TEXES has detected no spectral
features in BN at these wavelengths. The spectrograph slit was
roughly 1\farcs4 wide for all observations. The spectral coverage and
slit length varied depending on the setting. The slit was always long
enough that we nodded along the slit every 16 seconds to remove 
background emission.

The data were processed using the TEXES data reduction pipeline 
\citep{lacy_2002}.  
The pipeline operates on the raw data files and produces
optimally extracted, 1D spectra with a frequency scale set by user
identification of atmospheric lines. Using the difference of an
ambient temperature blackbody and the night sky emission, it is
possible to correct partially for telluric spectral features.  
Additional correction comes from comparison with the telluric
standard. The baseline has been removed by using fourth-order
polynomials.  
The range $1205-1227$~\cm{} shows
a ripple
whose period is roughly twice the observed linewidths
and that would severely compromise analysis of any line 
in this range.
Hence, this range has not been analyzed.

A realistic modeling of molecular lines requires an
estimation of the dust emission throughout the envelope.
It provides us with an estimation of the dust opacity
and temperature in different shells,
allowing us to determine how it affects the molecular emission.
Since the infrared continuum of CRL618 has remained 
roughly constant for several decades
\citep*[][C01a]{kleinmann_1978,pottasch_1984},
we have supplemented the TEXES/IRTF data with a
low resolution SWS/ISO spectrum (C01a) covering the range 
between 2.4 and 45~$\mu$m.
The properties of the dusty CSE derived from modeling
this continuum will be subsequently used to calculate the
synthetic profiles of the molecular lines 
(see \S\ref{sec:modeling}).

\section{Spectroscopic Data}
\label{sec:spectroscopy}

All the molecular species considered in this work
(\hcn, \acet, \diacet, \triacet, and
maybe \tetracet) are linear.
The frequencies and intensities of the lines of
\acet{} and \hcn{} have been extensively studied
in the laboratory but those of
\diacet, \triacet, and \tetracet{}
are still not accurately measured.
The frequencies of the ro-vibrational transitions of
the bands observed in our spectrum
are well known in the case of \diacet{} and \triacet.
However, the ro-vibrational spectrum of \tetracet{} 
remains unknown \citep{shindo_2001}.

\diacet{} is a 6-atom linear molecule with nine vibrational modes.
Four of them ($\nu_6-\nu_9$) are doubly-degenerated bending modes 
with energies of 625.4986, 482.7078, 627.8958 and 219.9778~\cm.
The rotational constant in the ground vibrational state
is 0.1464102~\cm{} \citep{guelachvili_1984}.
The spectroscopic constants have been taken from 
\citet{guelachvili_1984,mcnaughton_1992,arie_1992}, and 
\citet{matsumura_1982,matsumura_1984}.
The measured band intensity is
$\simeq 171$~\cmatm{} \citep{khlifi_1995}, implying that
the derived dipole moments of the ro-vibrational transitions 
of band $\nu_6+\nu_8$ are $\simeq 0.096$~D.

\triacet{} is a 8-atom molecule with 13 vibrational modes;
the last six ($\nu_8-\nu_{13}$) are doubly-degenerated bending modes.
Their energies are 622.38, 491.00, 258.00, 621.34, 443.50 and
105.04~\cm.
The rotational constant in the ground vibrational state is
0.044171~\cm.
All these data and the rest of the spectroscopic constants have been
taken from \citet{matsumura_1993,mcnaughton_1991}, 
and \citet{haas_1994}.
The dipole moments of the ro-vibrational transitions of band $\nu_8+\nu_{11}$
are $\simeq 0.12$~D, derived from a band intensity 
$\simeq 210$~\cmatm{} \citep{shindo_2003}.

The central frequency of the \tetracet{} 
$\nu_{10}+\nu_{14}$ band is not accurately known
but it is expected to fall
inside the observed range 
at $\simeq 1230$~\cm.
It has a small ground rotational constant,
$\simeq 2\times 10^{-2}$~\cm{} \citep{shindo_2001},
which will produce a band formed by a series of 
lines partially resolved with the resolution power provided by
TEXES, and a FWHM~$\simeq 5$~\cm{} at the temperatures prevailing in the
innermost CSE of CRL618.
This structure
could be easily identified if the abundance of this
species is large enough to produce absorption above 10\% of the
continuum, in view of the large density of lines from the other species.
The estimated dipole moment of this band is $\simeq 0.14$~D, 
which has been derived from the band intensity $\simeq 256$~\cmatm{}
\citep{shindo_2001}.

For these three molecules (\diacet, \triacet, and \tetracet{}), 
only the intensities of the whole bands 
are available in the literature.
It is worth noting that the observed bands are blended with several
hot bands (e.g., for \diacet{} the observed band is 
$\nu_6+\nu_8$ and the associated hot bands are $\nu_6+\nu_8+\nu_9-\nu_9$, 
$\nu_6+\nu_8+2\nu_9-2\nu_9$,\ldots) involving vibrational
states significantly populated at room temperature.
Therefore, we have estimated the
dipole moment of the ro-vibrational transitions of
the fundamental band of each molecule from the
intensity measurements quoted in the literature
and we have assumed that the hot bands 
($\nu_6+\nu_8+\nu_9-\nu_9$, $\nu_6+\nu_8+2\nu_9-2\nu_9$,\ldots)
have the same dipole moment as the $\nu_6+\nu_8$ one.
The consequences of this approximation will be discussed in 
\S\ref{sec:uncertainties}.

The frequencies and dipole moments of \acet{} and \hcn{} have been taken
from the HITRAN Database \citep{rothman_2005}.

Several bands from $^{13}$C-bearing isotopologues
of \hcn{} and the polyacetylenes are expected to fall in
the observed frequency ranges.
\citet{pardo_2007a} suggested a $^{12}$C/$^{13}$C ratio 
$\simeq 15$ in the torus, meaning that $[\hcn]/[\isohcn]\simeq 15$ and
$[\acet]/[\isoacet]\simeq [\diacet]/[\isodiacet]\simeq 
[\diacet]/[\isodosdiacet]\simeq 7.5$, where \isohcn, \isoacet,
\isodiacet, and \isodosdiacet{} are the most abundant
isotopologues after the main ones.
The \hcn{} lines present in the observed spectrum
are too weak to expect a detectable
absorption from \isohcn{} (Fig.~\ref{fig:f1}\textit{a}).
Concerning acetylene, although \isoacet{} lines might be assigned
to several very weak features (Fig.~\ref{fig:f1}\textit{a}),
their opacities are too low to affect the much stronger \acet{}
and \hcn{} lines or to be used to derive any reliable
information about \isoacet.
There is almost no information about the isotopologues of 
\diacet{} in the literature.
The center of the bands have been estimated from ab initio
calculations to be shifted between 5 and
10~cm$^{-1}$ from that of the main isotopologue \citep{mcnaughton_1992},
but these values could be affected by large uncertainties.
All the significant features in the frequency range $1228-1249$~\cm{} of
the observed spectrum can be explained
by considering just \diacet{} and \triacet{} (Fig.~\ref{fig:f1}\textit{b} 
and \textit{c}).
Therefore, either the bands of the isotopologues actually fall out of
the observed frequency range or their abundances are
insufficient to produce detectable lines.

\section{Description of the spectra and modeling}
\label{sec:modeling}

\begin{figure}
\centering
\includegraphics[angle=-90,width=0.45\textwidth]{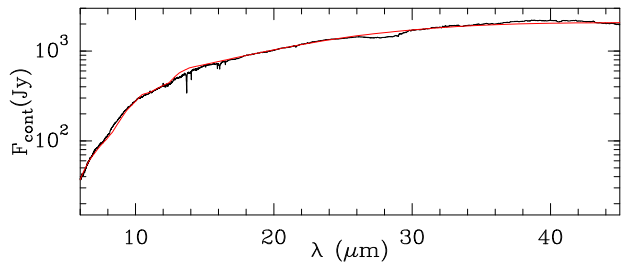}
\caption{Observed continuum of CRL618 (SWS/ISO; black) and fit between 6 and 
45~$\mu$m (red). A broad feature can be seen at $\simeq 28~\mu$m, which can be 
assigned to the solid state band produced by
the unknown material usually assumed to be MgS.
The fit has been calculated by assuming dust grains composed of
amorphous carbon (see text for details).}\label{fig:f2}
\end{figure}

The broad continuum of CRL618 taken with SWS/ISO displays
maximum emission at $\simeq 40~\mu$m,
typical of a cold envelope (see Fig.~\ref{fig:f2}).
The spectrum shows solid state bands, although much weaker 
than those displayed in the
continuum of the C-rich AGB star IRC+10216, where the bands due to
solid SiC at 11~$\mu$m and that at 27~$\mu$m 
\citep*[whose carrier
is still unknown but usually assigned to MgS; e.g.,][]{hony_2002}
are clearly 
present \citep{cernicharo_1999}.
Despite the complex geometry found in the
CSE of CRL618 (see \S\ref{sec:intro}), the broad 
continuum emission from
the dust grains smooths any geometrical effect in the spectrum.
Therefore,
the dusty AGB-CSE can be assumed to be spherically symmetric
in a first approximation.
We have used the model of the CSE developed by F08 and
assumed that the dust grains are composed of amorphous carbon 
\citep{rouleau_1991} to fit the continuum between 6 and 45~$\mu$m.
To date, this model cannot deal with radiative scattering by
dust grains. Hence, we limit our fitting to wavelengths above 6~$\mu$m.
The central source has been assumed to have an angular
size $\alpha_c\simeq 0\farcs135$ 
\citep{pardo_2004,pardo_2007b}, which was determined from 
a high-quality modeling of a large set of \hcccn{} lines in the
millimetric range.
We have adopted a distance to the source of 900~pc,
derived from 
the luminosity-distance relationship for PPNe and
estimations of the proper motion of dense inhomogeneities in 
the high velocity lobes
\citep{goodrich_1991,sanchezcontreras_2004a}.
These data imply a linear radius for the continuum source
$\rdin\simeq 1.8\times 10^{15}$~cm.
The CSE is assumed to have a radius of 600~$\alpha_c$,
since the model displays no significant contribution to the 
flux from gas and dust located at larger radii.
We have chosen a dust temperature following a $r^{-\gamma_d}$ law 
with different values for the exponent in the torus and the AGB-CSE.
The exponent $\gamma_d$ has been assumed to be $\simeq 0.39$ over the
latter region as in the case of 
IRC+10216 \citep*[F08;][]{sopka_1985}.
The synthetic spectrum has been fit to the data by eye, the
best fitting strategy when features such as solid state bands, which
cannot usually be well reproduced with current 
laboratory or theoretical data, are present.
The parameters derived following this procedure are shown 
in Table~\ref{tab:parameters}.
The largest differences between the synthetic and the observed spectra
come from the \acet{} and \hcn{} bands at 13~$\mu$m and the solid state
features between $20-40~\mu$m which are not considered in the fit
(Fig.~\ref{fig:f2}).
A discussion about the sensitivity of the synthetic spectrum to
variations in the parameters presented above can be found in 
\S\ref{sec:uncertainties}.

\begin{deluxetable}{cccc}
\tabletypesize{\footnotesize}
\tablecolumns{4}
\tablewidth{0.475\textwidth}
\tablecaption{Derived Parameters\label{tab:parameters}}
\tablehead{\colhead{Parameter} & \colhead{Units} & \colhead{Value} & \colhead{Errors}}
\startdata
N$(\hcn)$                        & cm$^{-2}$  & $\left(2.5\pm 0.9\right)\times 10^{17}$     & (6) \\
N$(\acet)$                       & cm$^{-2}$  & $\left(3.1\pm 0.7\right)\times 10^{17}$ & (2) \\
N$(\diacet)$                     & cm$^{-2}$  & $\left(2.1\pm 0.3\right)\times 10^{17}$ & (3,4) \\
N$(\triacet)$                    & cm$^{-2}$  & $\left(9.3\pm 1.4\right)\times 10^{16}$ & (5) \\
$v_\subscript{exp}(\rdin)$          & \kms      & $1.00^{+0.40}_{-0.22}$                     & (2)\\
$v_\subscript{exp}(\rmida)$          & \kms     & $8.00^{+0.20}_{-0.50}$                     & (2)\\
$v_\subscript{exp}(\rmidb)$          & \kms     & $17.0\pm 0.8$                           & (2)\\
\rout                             & \rdin    & $11.0\pm 1.0$                           & (1) \\
$T_\subscript{rot}(\hcn)$           & K        & $350\pm 50$                             & (6)\\
$T_\subscript{rot}(\acet)$          & K        & $200\pm 20$                             & (2)\\
$T_\subscript{rot}(\diacet)$        & K        & $100\pm 20$                             & (3)\\
$T_\subscript{rot}(\triacet)$       & K        & $100^{+100}_{-80}$                         & (5)\\
$T_c$                              & K         & $400\pm 50 $                          & (1) \\
\tdin                             & K         & $110\pm 6$                             & (1) \\
\tdout                             & K         & $98\pm 4$                             & (1) \\
$\tau_d$($8~\mu$m)                &           & $2.3\pm 0.3$                          & (1) \\
$\xi_\subscript{torus}$               & \%        & $65^{+12}_{-18}$                        & (1)\\
$\xi_\subscript{AGB}$                 & \%        & $35\pm 5$                             & (1)
\enddata
\tablecomments{
N($X$): total column density of species $X$;
$v_\subscript{exp}(r)$: gas expansion velocity at radius $r$;
$\rout$: radius of the outer boundary of the torus;
$\trot(X)$: rotational temperature of species $X$ at
the inner boundary of the torus;
$T_c$: black-body temperature of the central source;
$T_d(r)$: dust temperature at radius $r$;
$\tau_d$($8~\mu$m): dust optical depth at 8~$\mu$m;
$\xi_\subscript{torus}$ and $\xi_\subscript{AGB}$: fraction of the total dust 
optical depth coming from the torus and the AGB-CSE.
The uncertainties have been calculated by varying the parameters related to
(1)~continuum,
(2)~\acet{} $\nu_5$R$_e(21)$,
(3)~\diacet{} $\nu_6+\nu_8$R$_e(7)$,
(4)~\diacet{} $\nu_6+\nu_8$R$_e(22)$,
(5)~\triacet{} $\nu_8+\nu_{11}$R$_e(34)$, and
(6)~HCN $\nu_2$R$_e(22)$.
N(\acet) and N(\hcn) have been estimated
from the CSE and the HVG at once.
See \S\ref{sec:results} for separate estimations in each structure.
In the calculations, we have assumed the distance to
CRL618 and the size of the central source as fixed.
See the text (\S\ref{sec:uncertainties}) for a deeper
explanation about the strategy followed for the estimation of the
uncertainties.}
\end{deluxetable}

\begin{figure}
\centering
\includegraphics[width=0.45\textwidth]{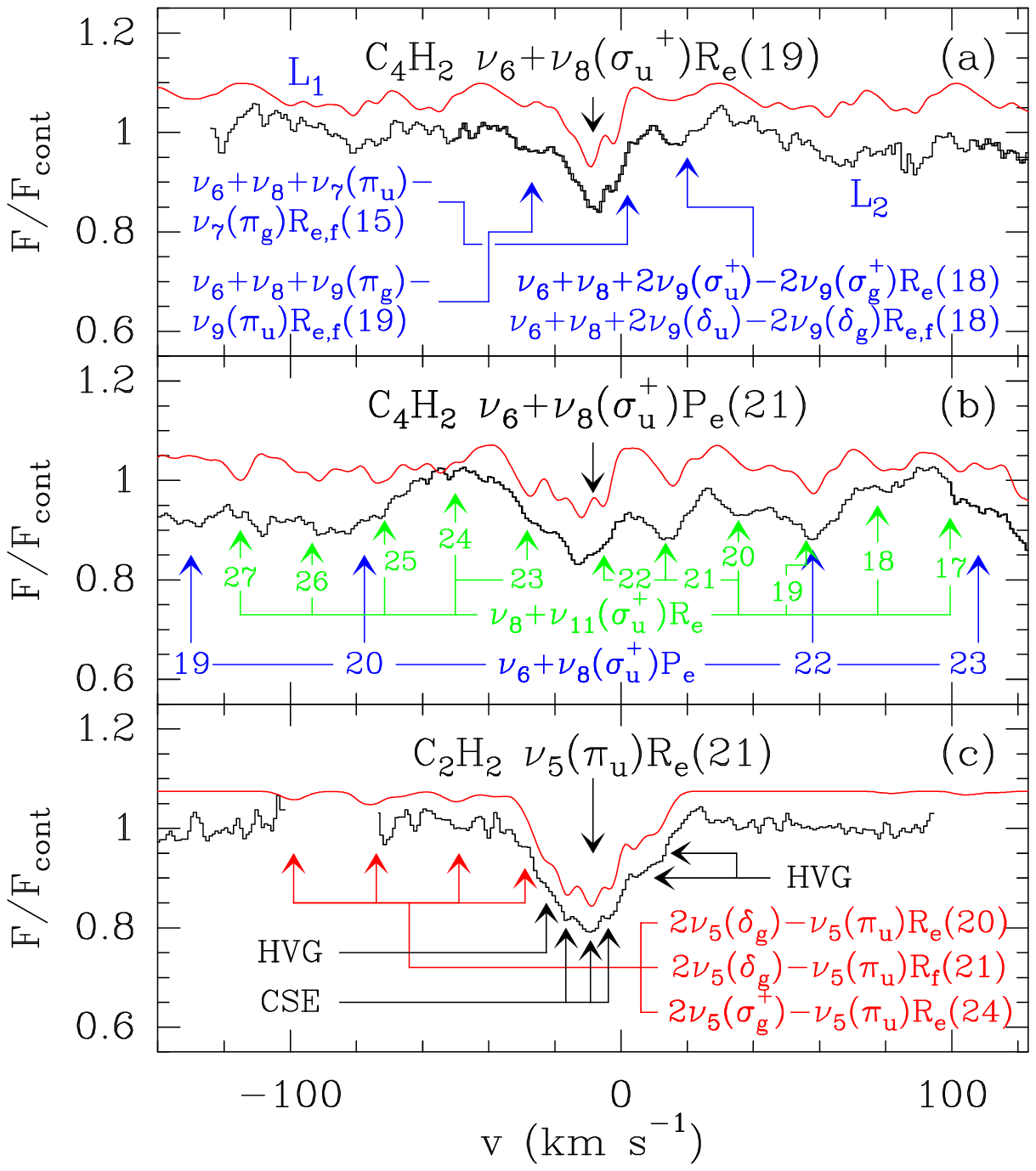}
\caption{Comparison between observed and synthetic \acet{} and \diacet{} lines.
The strongest lines are 
($a$) \diacet{} $\nu_6+\nu_8(\sigma_u^+)$R$_e(19)$,
($b$) \diacet{} $\nu_6+\nu_8(\sigma_u^+)$P$_e(21)$, and
($c$) \acet{} $\nu_5(\pi_u)$R$_e(21)$.
These spectra show additional \diacet{} and \triacet{} lines
(panel $a$:~$L_1$ and $L_2$ are \diacet{} 
$\nu_6+\nu_8(\sigma_u^+)$R$_e(20)$ and R$_e(18)$; 
panel $b$:~\diacet{} in blue and \triacet{} in green).
The velocity axis is centered in the LSR systemic velocity
\citep*[$\simeq -21.3$ \kms;][]{pardo_2004}.
The maximum absorptions in the strongest lines
arise from the CSE, displaying velocities
of $\simeq -3.5,~-9.0$, and $-16.0$~\kms.
The \acet{} line (panel $c$) shows the HVG contribution.
These features are not present in the \diacet{} lines implying that
they arise just from the CSE.
[\textit{See the electronic edition of the Journal for a color version of
this figure.}]}\label{fig:f3}
\end{figure}

We estimate the rms noise of the TEXES/IRTF spectrum to be $\sigma\simeq 1$\%.
With this noise level, we assume that a line is detected if its absorption
and/or emission is larger than $3\sigma$.
The part of the spectrum ranging from 778 to 784~\cm{} shows three \acet{} lines 
from band $\nu_5$ and two \hcn{} lines from band $\nu_2$.
However, there are some features in the spectrum that do not 
fulfil the $3\sigma$ condition but might be assigned to transitions of
the \acet{} bands $\nu_4+\nu_5-\nu_4$ and $2\nu_5-\nu_5$.
The strongest line in this range is \acet{} $\nu_5$R$_e(21)$.
The range between 1228 and 1249~\cm{} shows \diacet{} lines from the fundamental 
band $\nu_6+\nu_8$ and the hot bands
$\nu_6+\nu_8+\nu_9-\nu_9$, $\nu_6+\nu_8+2\nu_9-2\nu_9$, and (maybe) $\nu_6+\nu_8+\nu_7-\nu_7$ 
\citep*[ro-vibrational constants for these bands are from][]{mcnaughton_1992}.
The $\nu_8+\nu_{11}$ band of \triacet{} appears as weak lines often 
overlapped with those of
the \diacet{} $\nu_6+\nu_8$P$_e$ branch (Figs.~\ref{fig:f1} and \ref{fig:f3}).

Both \acet{} $\nu_5$ and \diacet{} $\nu_6+\nu_8$ lines
are broad ($\textnormal{FWHM}\simeq 35$ and $20$~\kms{}
with a spectral resolution of $\simeq 4$~\kms),
in contrast with those of the hot bands ($\simeq 10$~\kms; Fig.~\ref{fig:f3}).
The profiles of the lines of these fundamental bands exhibit a main absorption
surrounded by several features at lower and higher frequencies.
Some of these features are associated with hot bands and with \triacet{} lines from
the fundamental band $\nu_8+\nu_{11}$ (Fig.~\ref{fig:f3}).
\citet*[][hereafter P04]{pardo_2004} suggest a constant turbulence velocity 
in the torus $\simeq 3.5$~\kms{}
and an expansion velocity $\lesssim 18$~\kms{}, 
allowing many of the unassigned features within the line profiles 
to be attributed to \diacet{} and \acet{} expanding in this region. 
However, the 
origin of the remaining \acet{} line features is less certain since they require
expansion velocities significantly larger than that observed in the torus
(see Fig.~\ref{fig:f3}).
We estimate from our model that
the emission from the central source at 8~$\mu$m
is at least five orders of magnitude
larger than that coming from regions with an impact parameter $b>\alpha_c$.
Hence, these unexplained features could arise from gas in the HVG
located in front of the central source \citep*[P04;][]{sanchezcontreras_2004a}.

\begin{figure}
\centering
\includegraphics[angle=-90,width=0.47\textwidth]{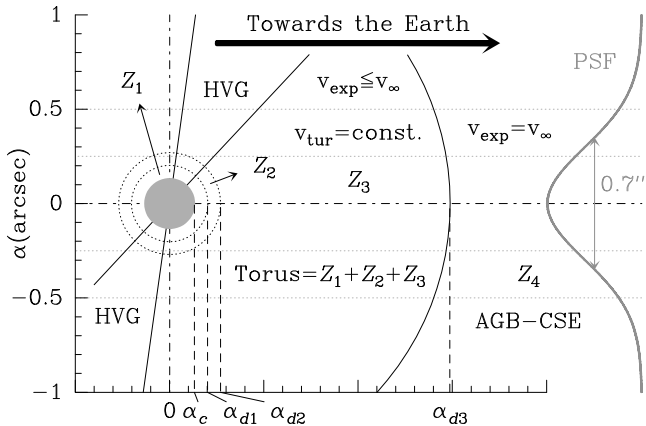}
\caption{Model of the inner envelope of CRL618.
The CSE is composed of four shells:
$\mathcal Z_1$, $\mathcal Z_2$, 
and $\mathcal Z_3$ inside of the torus, and $\mathcal Z_4$ 
corresponding to the AGB-CSE.
We have adopted an angular radius for the central source 
$\alpha_c\simeq 0\farcs135$ (P04).
The values of the angular 
radii $\alpha_{d1}$ and $\alpha_{d2}$ are assumed to be 
$1.5\alpha_c$ and
$2.0\alpha_c$, respectively, to allow the model to deal with unexpected 
sharp variations of the physical conditions.
The radius $\alpha_{d3}$ has been considered as a free parameter.
We have assumed a constant turbulent velocity 
$v_\subscript{tur}\simeq 3.5$~\kms{} in the torus (P04).
The high velocity gas (HVG) has been modeled as gas condensations
(see text).
The axis on the left shows the angular offset with
respect to the position of the source.
On the right, we have included the point spread funcion 
(PSF) of the telescope to allow comparisons with the model.}
\label{fig:f4}
\end{figure}

The model adopted to interpret the TEXES/IRTF
observations is a modified version
of that developed by F08. 
Although the innermost CSE displays roughly axial symmetry,
as several interferometric observations established
\citep{sanchezcontreras_2004a,sanchezcontreras_2004b}, we have assumed 
spherically symmetric
abundance distributions for the considered species
as a first approximation.
This is a reasonable approach
since the shells near the central source should not be 
significantly affected by the outflows.
The CSE has been divided in four regions ($\mathcal Z_1,\ldots,
\mathcal Z_4$) related to the torus and the AGB-CSE (see Fig.~\ref{fig:f4}).
We have assumed a gas density profile $\propto r^{-2}v_\subscript{exp}^{-1}$
(constant mass-loss rate), 
excitation temperature profiles (vibrational and rotational) 
$\propto r^{-\alpha}$, with $\alpha$ depending on the region,
and initially a constant turbulence velocity of 3.5~\kms{}
in the torus \citep{pardo_2004}.
We have added three gas condensations in the inner boundary
of region $\mathcal Z_1$ with different physical and chemical 
conditions with respect to the innermost CSE 
in order to model an inhomogeneous HVG.
The synthetic molecular lines have also been fit to the 
observed ones by eye.
This strategy allow us to fit lines partially overlapped with
unknown features or affected by irregular baselines impossible
to be removed (see Fig.~\ref{fig:f1}\textit{b}).
The derived parameters are shown in Table~\ref{tab:parameters}.

\subsection{Uncertainties}
\label{sec:uncertainties}

We estimated our uncertainties as in F08.
Specifically, we explore how the variation of a given
parameter affects either the continuum or a molecular
line, depending on the nature of the parameter.
When using a molecular line, we choose the one most 
sensitive to the variation of the parameter.
We consider a synthetic continuum/line as a good fit to the observed
one when all its points deviate from those of the best fit
in less than 10\% the maximum absorption when modeling 
a molecular line and the maximum emission when modeling the continuum.
Otherwise, the synthetic spectrum is considered as a bad fit.
The minimum and maximum values allowed for a given parameter
are the smaller and larger of its values which produce good fits when 
varying, at the same time, all the other parameters.
Therefore, the lower and upper uncertainties are the differences between
the minimum and maximum values and that derived
from the best fit.

In this work, we have fixed two parameters that cannot
be estimated from our observations
at the same time as the rest:
the distance to CRL618, $D$, and the angular radius of the central
source, $\alpha_c$.
To date, these parameters are not well known, specially $D$, 
which could be two times larger
(see \S\ref{sec:intro}).

A variation in any of these parameters affects the linear
radius of the central source.
An increase in the $\alpha_c D$ product would raise the optical
depth ($\tau_\nu\propto\alpha_c D$) diminishing
the escape probability of the infrared photons that come from the 
the inner shells of the torus.
This effect could be counteracted by decreasing the column
density of the considered molecular species and the dust
up to a factor $\simeq 2-3$ in our case.

In addition, a variation in $\alpha_c$ would have
an impact on the temperature profiles as well.
The most affected parameters would be the exponents while
the temperatures at the inner boundary of the torus
would require minor corrections to recover the best fit.

The approximations adopted to estimate the dipole moments
of the ro-vibrational transitions of \diacet, \triacet, and
\tetracet{} also introduce additional uncertainties in several
parameters.
The dipole moment of a ro-vibrational transition behaves as
a second degree polynomial on $J(J+1)$ 
\citep*[see, e.g.,][]{jacquemart_2001}, while we have
assumed that all the ro-vibrational transitions in a
band have the same dipole moment.
This different dependence on $J$ alters the
derived column densities and excitation 
temperatures from the real ones.
Adopting a maximum deviation of 10\% for the (constant)
dipole moment squared used in our model with respect to
the real one
\citep*[as occurs for \acet;][]{jacquemart_2001},
the derived column densities and excitation temperatures would 
be affected by an additional error up to 30\%.

Similar variations in the uncertainties of the vibrational
temperatures could occur due to having assumed no differences
between the vibrational dipole moments of the fundamental band
and the associated hot bands.

\section{Results and Discussion}
\label{sec:results}

The fits indicate that the deepest absorptions of all lines (Fig.~\ref{fig:f3})
have velocities $\simeq -3.5$, $-9.0$, and $-16.0$~\kms{}, and
come from regions $\mathcal Z_1$, $\mathcal Z_2$, and $\mathcal Z_3$, 
respectively.
This is compatible with an expansion velocity profile in which
$v_\subscript{exp}(\rdin)=1.0$, $v_\subscript{exp}(\rmida)=8.0$,
$v_\subscript{exp}(\rmidb)=17.0$, and $v_\subscript{exp}(r\gtrsim\rout)=18.5$~\kms{}.
The line \acet{} $\nu_5$R$_e(21)$ also displays several features
with velocities $\simeq 10.0$, $2.3$, and $-21.8$~\kms,
not present in the \diacet{} lines (Fig.~\ref{fig:f3}).
They do not arise from the CSE and are not hot bands.
This fact supports the existence of \acet{} in the HVG and suggests
that the abundance of \diacet{} is negligible in it.

The discrete absorptions found in the \acet{} lines
(Fig.~\ref{fig:f3}) produced by the HVG
could be the fingerprints of dense clumps
whose outermost layers protect the inner molecular gas
from being dissociated by the UV radiation field 
impinging on them and allow a significantly slower
chemical evolution in their cores \citep{redman_2003}.
In these circumstances, these clumps must be formed
due to ($i$) an irregular mass-loss in the latest
phases of the stellar atmosphere ejection \citep{dyson_1989}, or
($ii$) density fluctuations or instabilities in the 
ionization front 
\citep{capriotti_1973,garciasegura_1996,williams_1999}.
As the clumps must be in the HVG and in front of the central
source, we estimate that their characteristic size should
be $\lesssim 1~\rdin$.

We derive a vibrational temperature, \tvib, $\simeq 350$~K
for \acet{} in the innermost CSE and in the HVG. 
In the CSE, it decreases to $\simeq 200$~K
between \rdin{} and \rmida{} reaching $\simeq 50-100$~K at \rout{}. 
For \diacet{}, $\tvib\simeq 500$~K in the innermost CSE
followed by a fast decrease to $\simeq 400$~K at \rmida{} and reaching
$\simeq 200-250$~K at the end of region $\mathcal Z_3$.
It is not possible to derive a reliable \tvib{} for \triacet{} since its lines
are very weak but the fits support values similar to that of \diacet{}.
The \hcn{} lines do not allow an accurate \tvib{} determination in the
torus or in the HVG because of its low abundance.
However, the lack of absorption from the hot bands and emission from the
fundamental band implies $\tvib\lesssim 100$~K. 
The difference in \tvib{} between \hcn, \acet, and \diacet{} is likely produced by 
different pumping mechanisms between these species. 
In particular we note that some of the hot bands arise from 
metastable vibrational states (i.e., infrared-inactive).

The synthetic spectra show that the \acet{} and \diacet{} hot bands require
lower turbulence velocities ($\lesssim 2$~\kms) than fundamental bands 
($\simeq 3.5$~\kms). It suggests that the turbulence velocity would be
lower in regions $\mathcal Z_1$ and $\mathcal Z_2$, where most of the
hot bands arise, than in the rest of the torus.
We would need additional observations with a higher signal-to-noise ratio
and a more realistic model to explain this result.

Concerning the rotational temperatures, \trot, we derive $\trot\simeq 200$~K 
for \acet{}. This value has little error because
the observed lines are very sensitive to variations in \trot.
For \diacet, we derive $\trot\simeq 100$~K 
also accurately due to the large number of observed lines.
In the case of \hcn{}, $\trot\simeq 350$~K but this time
the uncertainties are large since we have only observed two weak lines.
The diversity of rotational temperatures indicate that these 
species are out of LTE, at least in regions $\mathcal Z_1$ and $\mathcal Z_2$,
which dominate the molecular absorption.

The results support a 3:1 ortho-para ratio for the 
polyacetylenes, i.e., the ratio expected under LTE conditions.
This fact implies that the formation of polyacetylenes
occurs in warm regions as suggested in previous works (e.g., C01a; C04),
where the ortho-para ratio of the new molecules is given by
the spin degeneracy.
Taking into account that the rotational contants of the polyacetylenes
are small compared to the kinetic temperatures prevailing in the
emitting regions, we do not expect ortho-para conversions even though
they are not strictly forbidden, contrarily to the case of H$_2$ at low temperatures
\citep{herzberg_1963}.
Moreover, our result also points out that other processes capable of producing
ortho-para conversions such as
($i$) proton exchange with atomic hydrogen, 
($ii$) collisions with paramagnetic molecules or ions such as
H$^+$, H$_2^+$, or H$_3^+$, and 
($iii$) spin inversion processes on the dust grain surfaces 
\citep{farkas_1935,herzberg_1963,massie_1982,burton_1992},
are inefficient in the innermost CSE.

The \acet{} column density in the CSE is $\simeq 2.0\times 10^{17}$~cm$^{-2}$
coming 45\% from region $\mathcal Z_1$,  
30\% from $\mathcal Z_2$, and 
25\% from $\mathcal Z_3$.
The contribution from the AGB-CSE (region $\mathcal Z_4$) is negligible.
The averaged column density in the HVG is $\simeq 1.1\times 10^{17}$~cm$^{-2}$.
Hence, the total column density is $\simeq 3.1\times 10^{17}$~cm$^{-2}$,
in good agreement with $2\times 10^{17}$~cm$^{-2}$, the value derived by 
C01a from their low spectral resolution SWS/ISO observations
(see Table~\ref{tab:parameters} for an error estimation).
In addition, the column density that we propose for the CSE agrees with
results of the
time-dependent chemical models developed by 
\citet{woods_2002,woods_2003}, which bracket it in the range 
$\simeq 5\times 10^{16}-5\times 10^{18}$~cm$^{-2}$,
depending on the evolution degree of the envelope.
It is possible to estimate the \acet{} abundance ratio between
the clumps expanding in the HVG and the gas in the innermost CSE
and to compare this result with those of \citet{redman_2003}.
Assuming a characteristic length in the line of sight
$\lesssim 0.5$~\rdin, the averaged gas density in the clumps is 
$\gtrsim 100$~cm$^{-3}$.
Hence, as the \acet{} density in the innermost CSE is 
$\simeq 300$~cm$^{-3}$, the ratio is $\gtrsim 0.3$.
Following the results by \citet{redman_2003}, this lower limit
in the ratio would be achieved after $\simeq 1100$~yr of evolution.
However, the age of CRL618 as a PPN has been estimated in several
hundreds of years \citep*[e.g.,][]{kwok_1984}.

The \diacet{} column density in the CSE is $\simeq 2.1\times 10^{17}$~cm$^{-2}$
(35\% from region $\mathcal Z_1$, 
60\% from $\mathcal Z_2$, and
5\% from $\mathcal Z_3$), again with insignificant
contribution from the AGB-CSE
(Table~\ref{tab:parameters}).
As we do not see \diacet{} in the HVG, the
total column density is that derived for the CSE which is
a factor $\simeq 2$ larger than $1.2\times 10^{17}$ cm$^{-2}$,
suggested by C01a.
The difference between both values is probably due to the
large PSF of ISO compared to that of IRTF
($\simeq 3\arcsec$ and $0\farcs7$ at $\simeq 8~\mu$m, respectively).
In this case, our result is very similar to $2.3\times 10^{17}$~cm$^{-2}$,
the maximum abundance for \diacet{} according to the model 
of W03 for the evolution of the CSE.
As the abundance of \diacet{} in the HVG
is much lower than in the innermost CSE,
we estimate the \diacet{} abundance ratio 
between the clumps and the innermost CSE in a factor $\lesssim 0.1$.
This upper limit supports the results by \citet{redman_2003},
which suggest a ratio $\simeq 0.01-0.3$ for clumps aged between
100 and 1000~yr.
This period of time and that derived from the \acet{} results
suggest that a faster chemistry could be
needed to better reproduce our results.
However, the discrepancies between their results and ours 
could have to do with differences in the gas density,
directly related to the adopted distance to the star, or
different simplifications regarding the detailed physical structure of PPN
\citep*[see more complex models in][]{pardo_2004,sanchezcontreras_2004a}.

For \triacet{}, the column density in the CSE is 
$\simeq 9.3\times 10^{16}$~cm$^{-2}$
(15\% from region $\mathcal Z_1$ and 85\% from $\mathcal Z_2$;
see Table~\ref{tab:parameters}).
There is no detectable contribution  to the observed \triacet{} lines
from region $\mathcal Z_3$ and from the HVG.
Hence, the total column density is in good agreement with
the value of $6\times 10^{16}$~cm$^{-2}$, derived by C01a.
The maximum column density proposed by W03
($\simeq 7.7\times 10^{16}$~\cm) is similar to our result but
somewhat lower.
It might indicate that the model developed by these
authors needs to be slightly improved as far as the interaction between
polyacetylenes and dust grains or unconsidered chemical reactions
involving polyacetylenes are concerned.

Finally, the column density of \hcn{} in the CSE is 
$\simeq 2.0\times 10^{17}$~cm$^{-2}$ and
the averaged column density in the HVG is $\simeq 5.0\times 10^{16}$~cm$^{-2}$.
Both results are affected by a large error
due to the weakness of the observed lines.
Hence, we can conclude that the total
column density is $\simeq 2.5\times 10^{17}$~cm$^{-2}$
(Table~\ref{tab:parameters}),
in good agreement with $1.5\times 10^{17}$~cm$^{-2}$
proposed by C01a but smaller than the value of 
$\simeq 4-7\times 10^{17}$~cm$^{-2}$, derived by \citet{pardo_2004,pardo_2005}.
The discrepancy existing with the latter result can be attributed
to \hcn{} molecules known to exist in the cold region $\mathcal Z_4$ (AGB-CSE),
unobserved through infrared observations.
As in the case of \diacet, the maximum column density expected by 
W03 is similar but larger than that derived by us, 
suggesting that their model works also fine for \hcn.

The abundance of \tetracet{} estimated from the results by 
C01a and that proposed by W03 and C04
suggests that the \tetracet{} $\nu_{10}+\nu_{14}$ band should be observed
in our spectrum.
Nevertheless,
we have not found any pattern of lines that could be assigned to
this species.
The telluric contributions to the raw spectra of CRL618 were removed 
by dividing the latter by the corresponding spectra of the BN object,
free of intrinsic molecular lines in the observed frequency range.
Further corrections were performed to mend minor instrumental features
from the spectra.
Therefore, we are sure we did not accidentally remove any
broad feature such as the \tetracet{} band.
The upper limit to its total column density,
calculated by assuming a peak intensity for the $\nu_{10}+\nu_{14}$
band lower than 10\% the continuum, and
the same excitation temperatures and coexisting
with \triacet,
is $5\times 10^{16}$~cm$^{-2}$.
% by adopting the same excitation temperatures and regions of
% existence than for \triacet{}.
This value is lower than the column densities of \acet{},
\diacet{} and \triacet{} in the CSE by a factor $4.0$, 
$4.2$ and $1.9$, respectively.

The abundance ratio $[\acet]/[\diacet]$ in the torus is $\simeq 0.95$,
compatible with a region where the abundances have reached
the steady state (C04).
The abundance ratios in each region ($\mathcal Z_1$, $\mathcal Z_2$, 
and $\mathcal Z_3$) are $1.2$, $0.49$, and $3.5$, respectively.
The abundances in $\mathcal Z_1$ seem to be in steady state while
in $\mathcal Z_3$ they are still evolving.
In region $\mathcal Z_2$ the ratio is very low indicating
variations in the abundances, likely related to
condensation onto dust grains
or involvement in chemical reactions still unconsidered
in the current chemical models.
The ratio $[\diacet]/[\triacet]\simeq 2.3$, in good agreement
with the results suggested by C04 for evolving abundances.
The \triacet{} lines are too weak to derive reliable ratios
for each region in the torus.

\subsection{SMP LMC 11, CRL618, and IRC+10216}

The protoplanetary nebula SMP LMC 11, located in the 
Large Magellanic Cloud,
is known to display several bands in absorption arising
from \acet, \diacet, and \triacet{}
\citep*[][and references therein]{bernardsalas_2006}.
This fact suggests that the chemistry in its CSE is very similar than
in CRL618.
Hence, a comparison between these two sources could contribute
interesting clues to a future chemical model of SMP LMC 11.

The observations carried out in the infrared by 
\citet{bernardsalas_2006} 
towards SMP LMC 11 show a wide absorption in the continuum
ranging from $12$ to $17~\mu$m, formed by the blending of 
several hundreds of lines of \acet, \diacet, \triacet, and
other abundant molecules such as C$_6$H$_6$.
From this feature can be inferred that \acet{} is more
abundant than in CRL618, even as much as in IRC+10216 
(with a column density $\simeq 1.6\times 10^{19}$~cm$^{-2}$; F08),
where a broad absorption
at $\simeq 13~\mu$m was also observed \citep{cernicharo_1999}.
The presence of strong hot bands such as $\nu_4+\nu_5-\nu_5$, 
$2\nu_5-\nu_5$ (at $13.5-14.0~\mu$m), and $\nu_4+\nu_5$
(at $\simeq 7.5~\mu$m) in absorption
indicates that the vibrational
temperature in the innermost CSE should be $\simeq 500-600$~K,
i.e., between those of CRL618
(with weak hot bands; see Fig.~\ref{fig:f1}(a)) and IRC+10216 
(with strong hot bands; F08).
This excitation would be probably caused by a somewhat
strong infrared radiation field produced by the warm dust,
as in the case of IRC+10216 (e.g., F08).

Concerning the rest of the polyacetylenes, the lower abundance ratio
between \triacet{} and \diacet{} in SMP LMC 11 than in CRL618
suggest than the former source is chemically less evolved than the
latter (W03,C04), in agreement with \citet{bernardsalas_2006}.
To date, it has been impossible to clearly detect \tetracet{} towards CRL618.
However, the great resemblance in their chemistry
and the good agreement between observations and chemical models
suggest that \tetracet{} could be built-up from \triacet{}
in the CSE of both sources.
\citet{bernardsalas_2006} compare in their Figure 2 two low
resolution mid-infrared spectra of CRL618 and SMP LMC 11.
These spectra seem to share a weak unidentified feature at
$\simeq 16~\mu$m, between the bands $\nu_8$ and $\nu_{11}$ 
of \diacet{} and \triacet, respectively.
This feature could be the Q branch of the \tetracet{}
band $\nu_{14}$, several times stronger than $\nu_{10}+\nu_{14}$
at room temperature \citep{shindo_2001}.
It should be observed in the future with higher 
spectral resolution to assess whether it can be assigned
to the $\nu_{14}$ band of \tetracet.
The recent discovery of C$_{60}$ and C$_{70}$ 
\citep{cami_2010}
strongly support the growth mechanism for carbon clusters 
proposed by C04.
In his models, carbon clusters where produced from the 
photodissociation of polyacetylenes
and, although limited to C$_{18}$, the predicted abundances for 
these large carbon species is very large.

\section{Summary and Conclusions}
\label{sec:conclusions}

In this Paper, we present high-resolution mid-infrared observations
towards the PPN CRL618.
The sampled spectral range ($778-784$ and $1227-1249$~\cm)
observed with the high resolving power spectrograph TEXES
allow us to resolve bands $\nu_6+\nu_8$ and
$\nu_8+\nu_{11}$ of \diacet{} and \triacet, respectively, 
in addition to several lines of bands $\nu_5$ and $\nu_2$
of \acet{} and \hcn.
These rich data have enabled the modeling of the useful
ro-vibrational line profiles of these molecular species to 
estimate their abundances and the physical conditions of the
gas and the dust throughout the inner circumstellar envelope.

The analysis of the observations has yield the following
results among others:
\begin{itemize}
\item our results support the chemical model suggested by W03 and C04 
for the polymerization of \acet;
\item most of the \hcn{} and \acet{} are in the inner
CSE. The rest come from several dense clumps located in the
high velocity gas;
\item \diacet{} and \triacet{} are formed in the innermost
CSE. Their abundances seem to be negligible in the clumps
since the emission from these
molecules is undetectable in our spectrum;
\item we are not able to detect any trace of the \tetracet{}
band $\nu_{10}+\nu_{14}$, expected to fall in the observed range.
This implies an even lower abundance for this species
compared to previously suggested
values. An upper limit to its column density has been estimated;
\item there exist large differences between the excitation
temperatures (vibrational and rotational) of \hcn, \acet, and
\diacet{} which indicate that the inner CSE is out of LTE.
\end{itemize}

In addition, the results of this work
demonstrate the power of IR observations in the
determination of the abundances and physical conditions of the gas
in complex structured environments such as the innermost envelopes of the 
evolved stars.
Further improvements in the search for \tetracet{} will be made
in the future by observing CRL618 and SMP LMC 11 at $\simeq 16~\mu$m
with the Echelon-cross-Echelle Spectrograph (EXES) mounted
on the Stratospheric
Observatory for Infrared Astronomy (SOFIA).

\acknowledgments

J. C. and J. P. F. would like to thank Spanish 
Ministerio de Educaci\'on y Ciencia for funding support through grant 
ESP2004-665, AYA2003-2785, and the ``Comunidad de Madrid''
government under PRICIT project S-0505/ESP-0237 (ASTROCAM). This study is
supported in part by the European Community's human potential
Programme under contract MCRTN-CT-2004-51230, ``Molecular Universe''.
During this study, J. P. F. was supported by the CSIC and the
``Fondo Social Europeo'' under internship grant from the I3P Programme,
by CONACyT under project SEP-2004-C01-47090, and by the UNAM through a
postdoctoral fellowship.
M. J. R. is supported by grant AST-0708074. TEXES was
built with funds from the NSF.
We would like to acknowledge the referee because of his/her useful
comments.

\end{document}